\documentclass[hidelinks,onefignum,onetabnum]{siamart250211}
\usepackage{lipsum}
\usepackage{amsfonts}
\usepackage{graphicx}
\usepackage{epstopdf}
\usepackage{algorithmic}
\usepackage{pifont}

\usepackage{amsmath,amssymb}
\usepackage{setspace}
\usepackage{lineno}
\usepackage{longtable}
\usepackage{caption}
\usepackage{makecell}
\usepackage{algorithm}
\usepackage{threeparttablex}
\usepackage{dsfont}
\usepackage[utf8]{inputenc}

\ifpdf
  \DeclareGraphicsExtensions{.eps,.pdf,.png,.jpg}
\else
  \DeclareGraphicsExtensions{.eps}
\fi

\newsiamremark{remark}{Remark}
\newsiamremark{hypothesis}{Hypothesis}
\crefname{hypothesis}{Hypothesis}{Hypotheses}
\newsiamthm{claim}{Claim}
\newsiamremark{fact}{Fact}
\crefname{fact}{Fact}{Facts}

\headers{
Probabilistic approach to longitudinal response prediction}{Isabella Cama, Michele Piana, Cristina Campi, Sara Garbarino}

\title{
Probabilistic approach to longitudinal response prediction: application to radiomics from brain cancer imaging
\thanks{\funding{CC acknowledges the financial support under the National Recovery and Resilience Plan (NRRP), Mission 4, Component 2, Investment 1.1, Call for tender No. 104 published on 2.2.2022 by the Italian Ministry of University and Research (MUR), funded by the European Union – NextGenerationEU– Project Title “Computational mEthods for Medical Imaging (CEMI) – CUP D53D23005830006 - Grant Assignment Decree No. 973 adopted on 30.6.2023 by the Italian Ministry of Ministry of University and Research (MUR). MP acknowledges the financial support of the "Hub Life Science – Digital Health (LSH-DH) PNC-E3-2022-23683267 - Progetto DHEAL-COM – CUP: D33C22001980001". Also, this research was supported in part by the MIUR Excellence Department Project awarded to Dipartimento di Matematica, Università di Genova, CUP D33C23001110001. IC, CC, MP, and SG are members of the Gruppo Nazionale per il Calcolo Scientifico - Istituto Nazionale di Alta Matematica (GNCS - INdAM).}}}

\author{
Isabella Cama \thanks{Universit\`a di Genova, Dipartimento di Matematica, via Dodecaneso 35, Genova, Italy, 16146}
\and Michele Piana\footnotemark[2] \thanks{IRCCS Ospedale Policlinico San Martino, Largo Rosanna Benzi 10, Genova, Italy, 16132}
\and Cristina Campi\footnotemark[2] \footnotemark[3]
\and Sara Garbarino\footnotemark[2] \footnotemark[3]}

\usepackage{amsopn}

\ifpdf
\hypersetup{
  pdftitle={aaa},
  pdfauthor={Isabella Cama, Michele Piana, Cristina Campi, Sara Garbarino}
}

\fi

\begin{document}

\maketitle

\begin{abstract}
\fontsize{9.5pt}{11pt}\selectfont
Longitudinal imaging analysis tracks disease progression and treatment response over time, providing dynamic insights into treatment efficacy and disease evolution. Radiomic features extracted from medical imaging can support the study of disease progression and facilitate longitudinal prediction of clinical outcomes. 
This study presents a probabilistic model for longitudinal response prediction, integrating baseline features with intermediate follow-ups. The probabilistic nature of the model naturally allows to handle the instrinsic uncertainty of the longitudinal prediction of disease progression. 
We evaluate the proposed model against state-of-the-art disease progression models in both a synthetic scenario and using a brain cancer dataset. Results demonstrate that the approach is competitive against existing methods while uniquely accounting for uncertainty and controlling the growth of problem dimensionality, eliminating the need for data from intermediate follow-ups. 
\end{abstract}


\begin{spacing}{1.2}   
\section{Introduction}

Longitudinal analysis of medical images enables the tracking of temporal changes, facilitating disease monitoring, treatment assessment, and outcome prediction. However, analyzing longitudinal images poses several challenges, such as managing variable imaging intervals and handling sparse data \cite{longitudinal_review}. 

Recent advances in literature indicate the potential for predicting longitudinal treatment outcome for various diseases using image-based features through machine learning (ML) techniques \cite{longitudinal_review}. These features, known as radiomic features, number in the thousands and can capture detailed information about morphology, intensity, and texture within a specific region of interest (ROI), such as a tumor. The standard radiomics workflow requires an initial step of image segmentation, relying on automatical techniques or manual delineation. Then, radiomic features are extracted from the segmented images and irrelevant or redundant features are discarded through feature selection procedures. Finally, the selected features are used to train predictive models for forecasting diagnosis, prognosis, or treatment response \cite{gillies2016radiomics}. 

The use of radiomics for longitudinal response prediction (or longitudinal outcome prediction) has been explored in literature for various applications. The primary approach to incorporate longitudinal information across timepoints includes feature concatenation over time. This method involves extracting radiomic features from multiple timepoints and combining them to construct a comprehensive radiomics-based representation of disease evolution \cite{longitudinal_tuberculosis,benitofarina}. This method enables to capture longitudinal changes in tissue characteristics that may not be evident from a single timepoint analysis. However, a significant drawback is the increased dimensionality of the model, which can introduce complexity and the risk of overfitting \cite{longitudinal_review}. 
Another approach is delta radiomics, that quantifies the relative change in radiomic features between different imaging timepoints, providing a characterization of tumor alterations over time \cite{longitudinal_breast_mri_delta,delta_longitudinal_hepatocellular}. This method is typically applied to data from two consecutive timepoints, but some studies extend this approach to three or more \cite{longitudinal_review}, where pairwise delta features are computed between timepoints and combined in different ways within predictive models \cite{delta_three_timepoints}. 
These methods are commonly employed to predict treatment response or disease progression across various medical conditions, such as osteoporosis \cite{longitudinal_bone}, tuberculosis \cite{longitudinal_tuberculosis}, breast cancer \cite{breast_cancer_longitudinal_concatenation, longitudinal_breast_mri_delta}, brain cancer \cite{longitudinal_brain_metastasis, longitudinal_delta_head_and_neck}, non-small cell lung cancer \cite{nsclc_longitudinal, benitofarina}, hepatocellular carcinoma \cite{delta_longitudinal_hepatocellular}, and neurodegenerative diseases \cite{delta_parkinson_longitudinal}, using radiomics from CT, PET, and MR imaging.
Nevertheless, both feature concatenation and delta radiomics face challenges in determining optimal regions for radiomic feature extraction at subsequent timepoints, for example after tumor resection, making it difficult to accurately characterize disease evolution \cite{longitudinal_review}. Additionally, missing data across longitudinal timepoints due to dropout can limit the applicability of such models for longitudinal disease characterization \cite{longitudinal_review}. For these reasons, some studies rely exclusively on pre-operative or baseline radiomics to forecast successive follow-ups \cite{longitudinal_bone}. However, this last approach does not fully leverage the longitudinal characteristics of the data. 

Building upon this foundation, we introduce a method that utilizes baseline radiomics but also additionally incorporates a probabilistic framework to forecast future response that quantifies uncertainty of intermediate outcomes, eliminating the need for radiomics data from intermediate follow-ups, and ultimately enhancing the longitudinal prediction. 

We validate our model on both a synthetic longitudinal dataset and radiomics data from glioblastoma patients \cite{lumiere}. Glioblastoma is one of the most aggressive primary brain neoplasms, with high predisposition to recurrence \cite{glioblastoma1, glioblastoma2}, making it an interesting application for a longitudinal study. Early detection of disease progression can facilitate timely therapeutic intervention that may consequently delay the relapse \cite{glioblastoma_longitudinal_intro}.
Numerical results show that our probabilistic model for prediction of longitudinal response performs similarly to SOTA references \cite{benchmark_glioblastoma}, consistently improving the longitudinal prediction solely based on baseline features, even with reduced dataset size.

By developing this longitudinal model, our work aims at advancing longitudinal radiomic analysis from a methodological perspective. Moreover, this approach might have potential orienting treatment decisions based on a dynamic understanding of a patient’s disease profile, and can be generalizable beyond oncology and radiomics to other longitudinal applications. \\

The paper is organized as follows. Section \ref{sec:longitudinal_approach} details the proposed probabilistic model for longitudinal response prediction. Section \ref{sec:data} describes the custom-designed synthetic longitudinal dataset and the real-world dataset \cite{lumiere}. Sections \ref{sec:benchmark} outlines the benchmark experiments for comparison with our method, while Section \ref{sec:ml_models} provides details on the machine learning models trained for this study. Section \ref{sec:results_longitudinal} presents the numerical results of the experiments on longitudinal prediction, providing a comparative analysis of the experimental findings. The conclusions follow in Section \ref{sec:conclusions_longitudinal}.

\section{Probabilistic approach to longitudinal response prediction} \label{sec:longitudinal_approach}

This section presents a novel, probabilistic approach to longitudinal response prediction, integrating baseline features with a probabilistic estimate of intermediate responses, to forecast the response at successive follow-ups. The longitudinal method we propose is applicable to any number of timepoints. However, for simplicity and without loss of generality, in the following we refer specifically to the first and second follow-ups after the baseline timepoint. In the context of predicting glioblastoma response to treatment, baseline features can be interpreted as pre-operative radiomic features, first follow-up features as those extracted from the first post-operative imaging, while the response at the first and second follow-up refer to the response to treatment at the first and second post-operative follow-up, respectively. \\

Let $\mathcal{C} = \{0,1\}$ be the target label space, and $\mathcal{F} \subset \mathbb{R}^{M}$ be the space of baseline features, extracted at the baseline timepoint, with $M$ number of features. The subscript $i$ indicates the $i$-th train-test data split. 

For the $i$-th data split, $f_{1}^i$ $f_{2}^i:\mathcal{F} \longrightarrow \mathcal{C}$ are the ML models to predict the response outcome at the first and second timepoint after baseline (first and second follow-up), respectively, using baseline features, and $i = 1,..., \# \text{splits}$. These models, $f_{1}^i$ and $f_{2}^i$, will be referred to as \textit{baseline models}. 
For the $j$-th patient of the $i$-th data split, $x^j \in \mathcal{F}$, $f_{1}^i(x^j) = \texttt{proba}_1^{i,j}$ and $f_{2}^i(x^j) = \texttt{proba}_2^{i,j}$ are the probabilities of belonging to class $1$ outputted by the models, while  $y_1^{i,j} = \mathds{1}_{\{\texttt{proba}^{i,j}_1 \geq 0.5\}}$ and $y_2^{i,j} = \mathds{1}_{\{\texttt{proba}^{i,j}_2 \geq 0.5\}}$ are the predicted response labels at the first and second follow-up, with $y_1^{i,j}$, $y_2^{i,j} \in \mathcal{C}$. 

		\begin{algorithm}[h!]
		\caption{Probabilistic modeling of longitudinal outcome \label{longitudinal}}
		{\bf{Step $\boldsymbol{0}$:}} \textbf{extract baseline features} 

		{\bf{Step $\boldsymbol{1}$: baseline model for predicting response at 1st follow-up}} \\
		For $i$ data split (train-test):
		\begin{enumerate}
			\item train classifier $f_{1}^i$ 
				\begin{itemize}
					\item with baseline features
					\item to predict response at $1$st follow-up, $y_{1,true}^{i,j}$
				\end{itemize}
		
			\item for each patient $j$ \\ 
			the predicted response is obtained by
			\begin{itemize}
				\item $f_{1}^i(x^j) = \texttt{proba}_1^{i,j}$
				\item $y_1^{i,j} = \mathds{1}_{\{\texttt{proba}^{i,j}_1 \geq 0.5\}}$
			\end{itemize}
			\item$\forall j$ patient in the test set
				\begin{itemize}
				\item record probability of response 1
				 at $1$st follow-up, $\texttt{proba}_1^{i,j}$
			\end{itemize}
		\end{enumerate}

		{\bf{Step $\boldsymbol{2}$: modeling probability distribution of response label 1 at $\boldsymbol{1}$st follow-up}}  \\
		For each patient $j$:
		\begin{enumerate}
			\item collect the $\texttt{proba}_1^{i,j}$ across splits from step $1$
			\item use G-KDE to fit the probabilities $\{\texttt{proba}^{i,j}_1\}_{i\text{-th split}}$ to obtain $\hat{Y}_1^{j}$
			\item sample $\widehat{\texttt{proba}}_1^j$ from $\hat{Y}_1^{j}$ $100$ times $\rightarrow \{\widehat{\texttt{proba}}_1^{j}\}_{k=1...100}$
			\item  $\hat{y}_1^{j} = \mathds{1}_{\{\widehat{\texttt{proba}}^{j}_1 \geq 0.5\}}$ is the $k$-th sampled response at $1$st follow-up for patient $j$
		\end{enumerate}
		{\bf{Step $\boldsymbol{3}$: longitudinal model for predicting response at 2nd follow-up}} \\
		For $i$ data split:
			\begin{enumerate}
			\item train classifier $f_{2}^{L,i}$ 
			\begin{itemize}
				\item with baseline features and $y_{1,true}^{i,j}$
				\item to predict response at $2$nd follow-up, $y_{2,true}^{i,j}$
			\end{itemize}
			\item for $k=1:100$:\\
			\text{          }for $j$ patient in the test set:
			\begin{itemize}
				\item $y_2^{L,i,j} = f_{2}^{L,i}(x^j, \hat{y}_1^{j})$ is the longitudinal prediction of treatment response at $2$nd follow-up

			\end{itemize}
            \vspace{0.2cm}
			\text{ } compute skill scores for $k$-th sample
			\item average skill scores across all samples\\
			
		\end{enumerate}

	\end{algorithm}

	An additional model is designed to predict the response at the second follow-up, using the combination of baseline features and the response at the first follow-up, as input feature, to explicitly account for the intermediate outcome in the training phase. For the $i$-th data split, $f_{2}^{L, i}:\mathcal{F} \times \mathcal{C} \longrightarrow \mathcal{C}$ constitutes the \textit{longitudinal model}. 

In the experimental setup, the true response labels at intermediate follow-up are available, allowing the model $f_{2}^{L, i}$ to be tested using baseline features and actual responses $y_{1, true}^{i,j}$. 
However, this is not the case for prospective studies, in which only baseline features are available (e.g., for newly enrolled patients who only have pre-operative radiomics from their scans), or in case of missing data. If the intermediate response is assumed to be unavailable during the model's testing phase, modeling the feature representing the intermediate response becomes crucial to benefit from the longitudinality of the model itself. In this work, the feature 'response at $1$st follow-up' is modeled in the following ways:
\begin{itemize}
	\item by using the predicted label $y_1^{i,j}$;
	\item by using the label $\hat{y}_1^{j}$, obtained by sampling from the probability distribution $\hat{Y}_1^j$ and thresholding, where $\hat{Y}_1^j$ is an approximation of the ideal probability distribution $Y_1^j$ of presenting response of class $1$ (e.g., progressive disease) at the first follow-up for each patient. 
\end{itemize}
The last option allows to incorporate in $f_{2}^{L, i}$ the uncertainty of the prediction outcomes at the first follow-up, i.e., the intermediate timepoint, derived from the ML model $f_{1}^i$.

	To approximate the distribution $Y_1^j$, we leverage the prediction outcome of the models $f_{1}^i$ across multiple data splits. Whenever a patient $x^j$ from the $i$-th split is included in the test set, we record their probability $\texttt{proba}_1^{i,j}$ of presenting class 1 response at the first follow-up, as predicted by the current trained model $f_{1}^{i}$. This procedure is repeated across all splits. As a result, each patient has a number of probabilities corresponding to the number of times they appear in the test set of a split. Then, for each patient, Gaussian Kernel Density Estimation (G-KDE) \cite{gkde, silverman2018density} is used to model the distribution $Y_1^j$ by fitting the probabilities $\{\texttt{proba}^{i,j}_1\}_{i\text{-th data split}}$, thus obtaining $\hat{Y}_1^{j}$. 
	For testing the longitudinal model $f_{2}^{L,i}$, $\widehat{\texttt{proba}}_1^j$ is sampled from $\hat{Y}_1^{j}$ one hundred times, the probabilities are thresholded as $\hat{y}_1^{j} = \mathds{1}_{\{\widehat{\texttt{proba}}^{j}_1 \geq 0.5\}}$, and the response labels at first follow-up $\hat{y}_1^{j}$ are assigned to each patient. For the $j$-th patient of the $i$-th data split, $y_2^{L,i,j} = f_{2}^{L,i}(x^j, \hat{y}_1^{j})$ is the longitudinal prediction of the response at the second follow-up. In this way, the longitudinal model is tested one hundred times, one for each sampling, and the prediction results can be averaged. 
    A step-by-step explanation of our probabilistic approach to longitudinal prediction of (treatment) response is reported in Algorithm \ref{longitudinal}.

\section{Data}\label{sec:data}
    To evaluate the performance of the longitudinal model introduced in Section \ref{sec:longitudinal_approach}, based on the probabilistic modeling of intermediate treatment response, we apply the described method to a synthetic tabular dataset consisting of three timepoints - baseline, first, and second follow-up - as well as to the Lumiere dataset, a publicly available radiomic dataset of glioblastoma progression \cite{lumiere}.

\subsection{Radiomic features}\label{sec:radiomics}
    Radiomic features are quantitative descriptors extracted from medical images that provide information about the shape, texture, intensity, and other characteristics of a ROI, such as a tumor or an organ. These features help analyzing the heterogeneity of tissues and can be used for disease diagnosis, prognosis, and treatment response prediction.
    
    Shape-based features describe the geometric properties of a region of interest, such as volume, surface area, and compactness. First-order statistics capture the distribution of voxel intensities within the image, summarizing global intensity characteristics. Gray Level Co-occurrence Matrix (GLCM) features assess spatial relationships between intensity values, reflecting textural uniformity and contrast. Gray Level Run Length Matrix (GLRLM) features quantify the length and distribution of consecutive voxel intensity runs, characterizing texture smoothness. Gray Level Size Zone Matrix (GLSZM) features evaluate the size and distribution of homogeneous intensity zones. Neighbouring Gray Tone Difference Matrix (NGTDM) features measure texture strength and contrast based on intensity differences with neighboring voxels. Gray Level Dependence Matrix (GLDM) features capture the degree to which voxels depend on neighboring intensities, quantifying image granularity and texture complexity. Extensive documentation of radiomic features can be found in \cite{pyradiomics}.

\subsubsection{Lumiere longitudinal dataset}
\begin{figure}[t]
		\includegraphics[width=\linewidth]{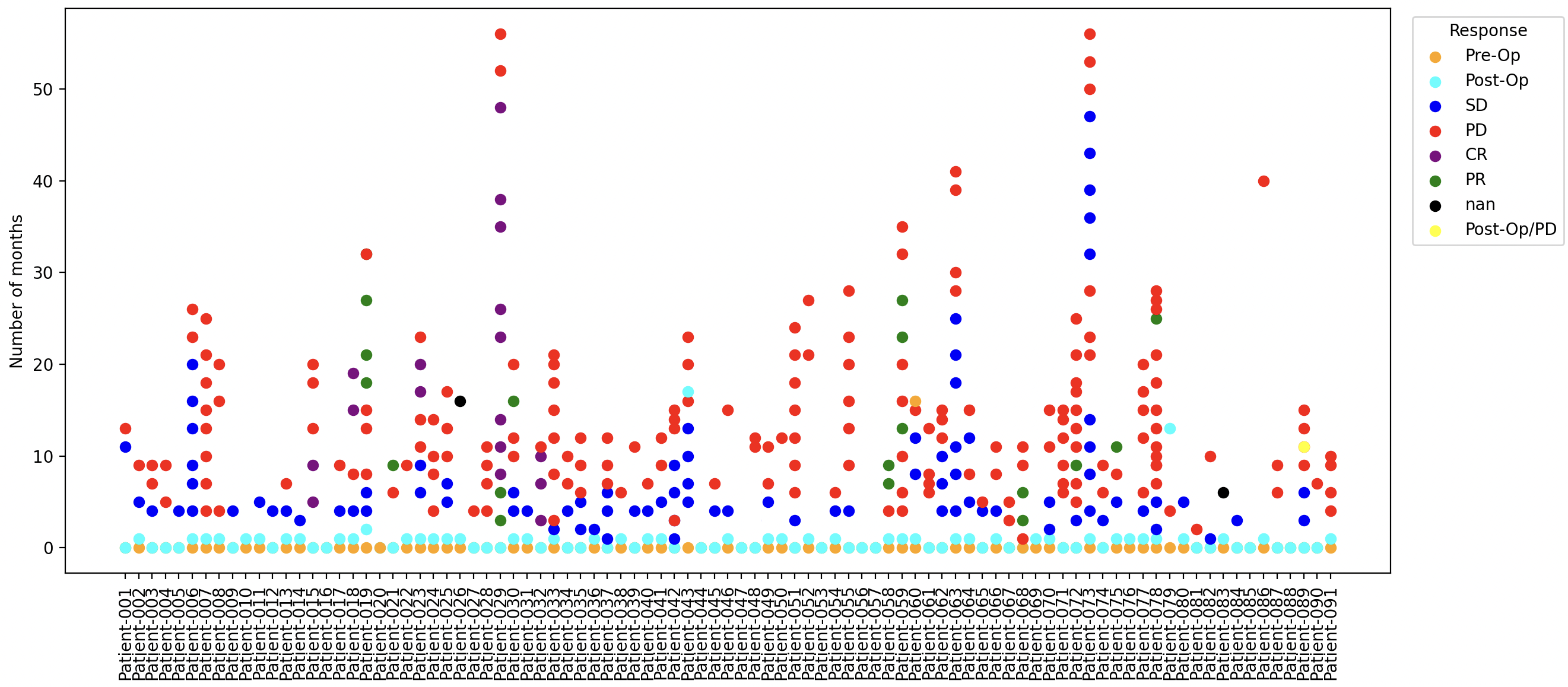}
		\caption[Lumiere dataset overview.]{All the patients included in the Lumiere dataset, with the response assessment at each timepoint. Pre-Op and Post-Op indicate pre- and post-operative timepoint; SD stable disease; PD progressive disease; CR complete response; PR partial response; nan indicates missing information about the timepoint; Post-Op/PD indicates second post-operative progression.}
		\label{fig:lumiere}
	\end{figure}

	Lumiere dataset \cite{lumiere} is a single-center longitudinal glioblastoma MRI dataset of selected follow-up studies. It includes expert ratings of response to treatment, performed according to the Response Assessment in Neuro-Oncology criteria (RANO)\cite{rano}: progressive disease (PD), stable disease (SD), partial response (PR), or complete response (CR). This collection includes MRI data of $91$ glioblastoma patients with a total of $638$ study dates and $2487$ images (age at surgery: mean $62.4$ years, std $10.3$ years, min $39.7$ years, max $80.4$ years; sex: $44$ female, $47$ male), having different and follow-up intervals (overall survival: mean $589.1$ days, std $334.0$ days, min $25$ days, max $1412$ days; time to progression: mean $305.4$ days, std $266.1$ days, min $25$ days, max $1537$ days).  Figure \ref{fig:lumiere} illustrates all the patients included in Lumiere dataset, along with the response assessment at each timepoint across available follow-ups.
	For a subset of patients, pathology information regarding MGMT methylation ($11$ not available) and IDH$1$ are present ($23$ not available). The dataset includes T$1$-weighted pre- and post-contrast, T$2$-weighted, and FLAIR MRI. Segmentations obtained through the automated segmentation tools DeepBraTumIA (\url{https://www.nitrc.org/projects/deepbratumia}) and HD-GLIO-AUTO \cite{hdglio, nnunet} (\url{https://github.com/CCI-Bonn/HD-GLIO}) are included, both based on U-Net deep learning architecture \cite{unet}. 
	Both methods include a co-registration of the four MRI sequences. HD-GLIO-AUTO registers all sequences to a reference image chosen automatically during processing, and DeepBraTumIA registers to an atlas such that all studies may be analyzed in the same space. 
	HD-GLIO-AUTO provides segmentations for the contrast-enhancing tumor and the T$2$-signal abnormality. DeepBraTumIA outputs labels for necrosis, contrast enhancement, and edema. For this study, DeepBraTumIA segmentations were used.
	
	Subject to data availability, radiomic features are provided by Lumiere dataset, extracted from the co-registered and resampled images for each segmentation label (PyRadiomics version 3.0.1). Z-score normalization, scaling by a factor of $100$, and intensity shifting by $300$ are used to harmonize the value range. Feature types include first-order statistics, $3$D shape, GLCM, GLRLM, GLSZM, NGTDM, and GLDM features, for a total of $110$ features for each mask label. No image filtering is considered for feature extraction. \\

	To obtain a homogeneous longitudinal dataset suitable for testing our methodology, we curated Lumiere dataset to include patients with a minimum of three timepoints (baseline, first follow-up, and second follow-up). Patients who exhibited complete or partial treatment responses were excluded, as their numerosity across all timepoints was insufficient to support training a four-class model (CR vs. PR. vs. PD vs. SD). Consequently, our experiments focused on distinguishing between stable disease (SD = $0$) and progressive disease (PD = $1$). 
	Given the variability in follow-up intervals among patients in the Lumiere dataset, we introduced an inclusion criterion based on the maximum time from baseline (pre-operative stage) to the first and second post-operative follow-up. Specifically, the first follow-up was required to occur within $20$ months from the baseline timepoint, and the second follow-up within $20$ months from the first. Patients whose first follow-up occurred more than $20$ months after surgery, or whose second follow-up exceeded $20$ months after the first, were excluded from the analysis.

\begin{figure}[t!]
		\includegraphics[width=\linewidth]{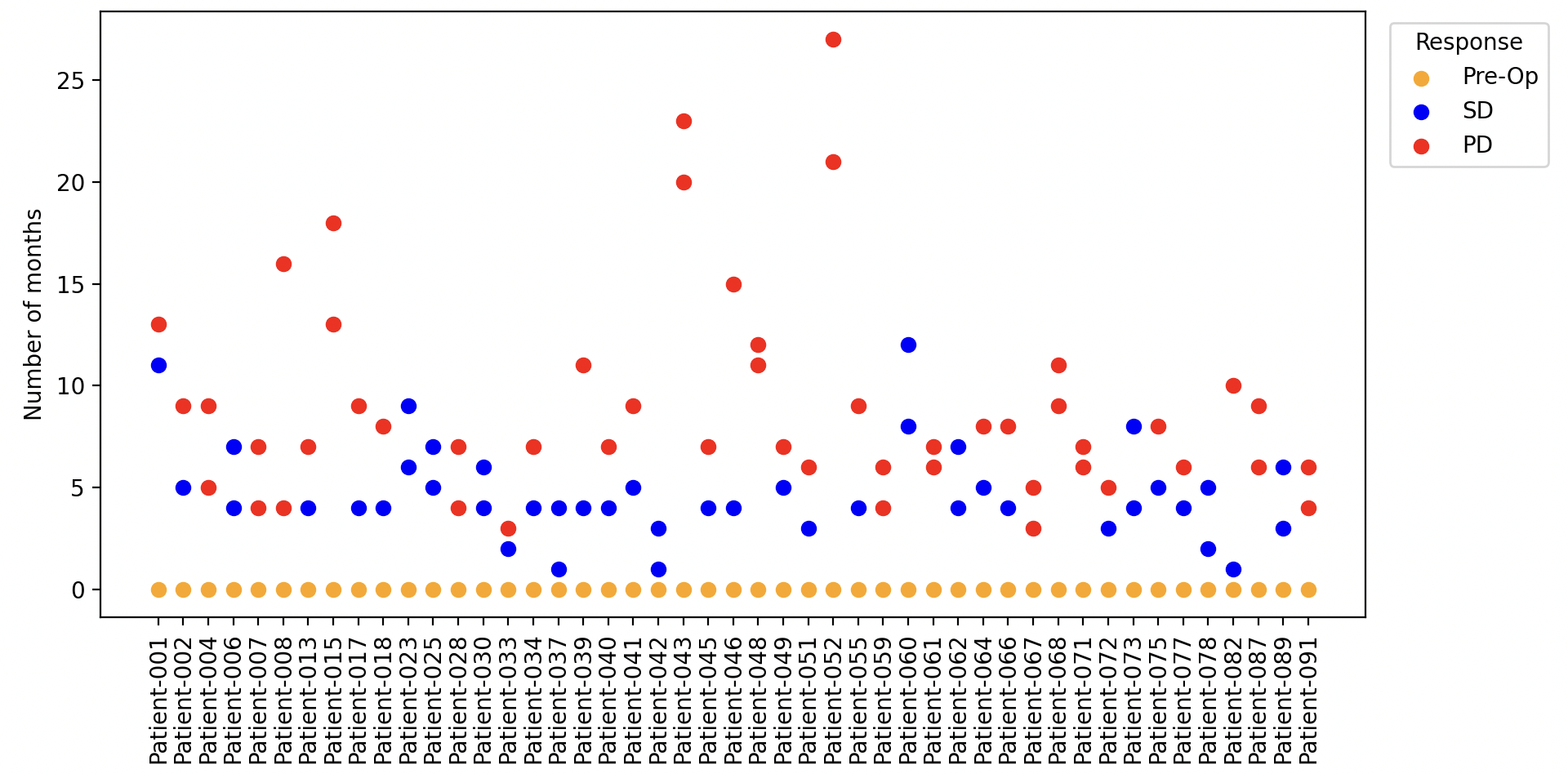}
		\caption[Eligible patients of Lumiere dataset.]{Subset of Lumiere dataset meeting the inclusion criteria, with response assessment at the first and second follow-up after baseline (pre-op label in the legend). For the benchmark experiment based on radiomics at the first follow-up, exclude patients $2$, $7$, $23$, $30$, $39$, $60$, $62$, and $89$.}
		\label{fig:lumiere_subset}
	\end{figure}

	For training our longitudinal model we required radiomics from all segmentation labels, i.e., contrast enhancement, necrosis, and edema, at the baseline timepoint on T$1$-weighted imaging. Following this selection, $47$ patients were deemed eligible. Additional selection was applied for the benchmark experiments, described in Section \ref{sec:benchmark}, involving first follow-up radiomics, requiring radiomics from the complete set of segmentation labels on T$1$ images at the first follow-up. After this step, $38$ patients remained eligible for the analyses. Figure \ref{fig:lumiere_subset} displays patients meeting the inclusion criteria.

\subsubsection{Synthetic longitudinal dataset}
Following the structure of Lumiere longitudinal dataset, we construct a synthetic version that preserves the same proportions of outcome labels at the first and second timpoints (stable vs. progressive disease). Our objective is not only to replicate the outcome distribution but also to ensure that the generated features follow a coherent logic. Specifically, we aim to create a longitudinal dataset where we know a feature-based separation is possible, in contrast to the real-world scenario, where such separation may not necessarily exist, or be very difficult to achieve, and use it to validate the our methology for longitudinal prediction in a simulated setting.
With this objective, we used a linear combination of baseline features to simulate the probability of progression at the first follow-up timepoint. Subsequently, we simulate a second follow-up by applying the transition probabilities derived from the Lumiere dataset, calculated based on the observed frequencies of transitions from stable to progressive and vice versa.

To construct the synthetic longitudinal dataset, we utilized existing radiomic features extracted from the original T$1$ images of the Lumiere dataset, randomly chosen among the available features: shape Surface-Volume Ratio ($x_1$), first order Median ($x_2$), first order Entropy ($x_3$), glcm Difference Entropy ($x_4$), and glcm Maximum Probability ($x_5$). Such features were considered as baseline features for the synthetic longitudinal dataset. To construct the first follow-up response, we followed this procedure. First, we drew the distribution of these features using the Kernel Density Estimation (G-KDE) method  \cite{gkde, silverman2018density} and sampled $300$ data points from the distributions. Next, we assigned a coefficient to each feature. The vector of coefficients is denoted as $\boldsymbol{\alpha}$.
The features were then standardized within the range $[-1, 1]$, therefore $\boldsymbol{x} = [x_1, x_2, x_3, x_4, x_5]$, $\boldsymbol{x} \in [-1,1]^{300 \times 5}$.
Then, by multiplying the matrix $\boldsymbol{x}$ with $\boldsymbol{\alpha}^T$, we obtained a vector $\boldsymbol{p} \in \mathbb{R}^{300\times 1}$ containing the scoring of each response at the first follow-up and representing the likelihood of having progressive disease. These scores were then converted into a binary column with values \(0\) and \(1\), based on whether their value was non-negative (assigned \(1\)) or negative (assigned \(0\)). The obtained labels represented the outcome of the first timepoint of the synthetic dataset. We chose $\boldsymbol{\alpha} = [0.4, 0.8, 0.5, 0.7, 0.2]$ through a trial-and-error approach, ensuring that the linear combination of the selected features would result in balanced outcome labels at the first follow-up, according to the outcome proportions provided by Lumiere dataset at the first follow-up.

\begin{figure}[t!]
	\includegraphics[width=\linewidth]{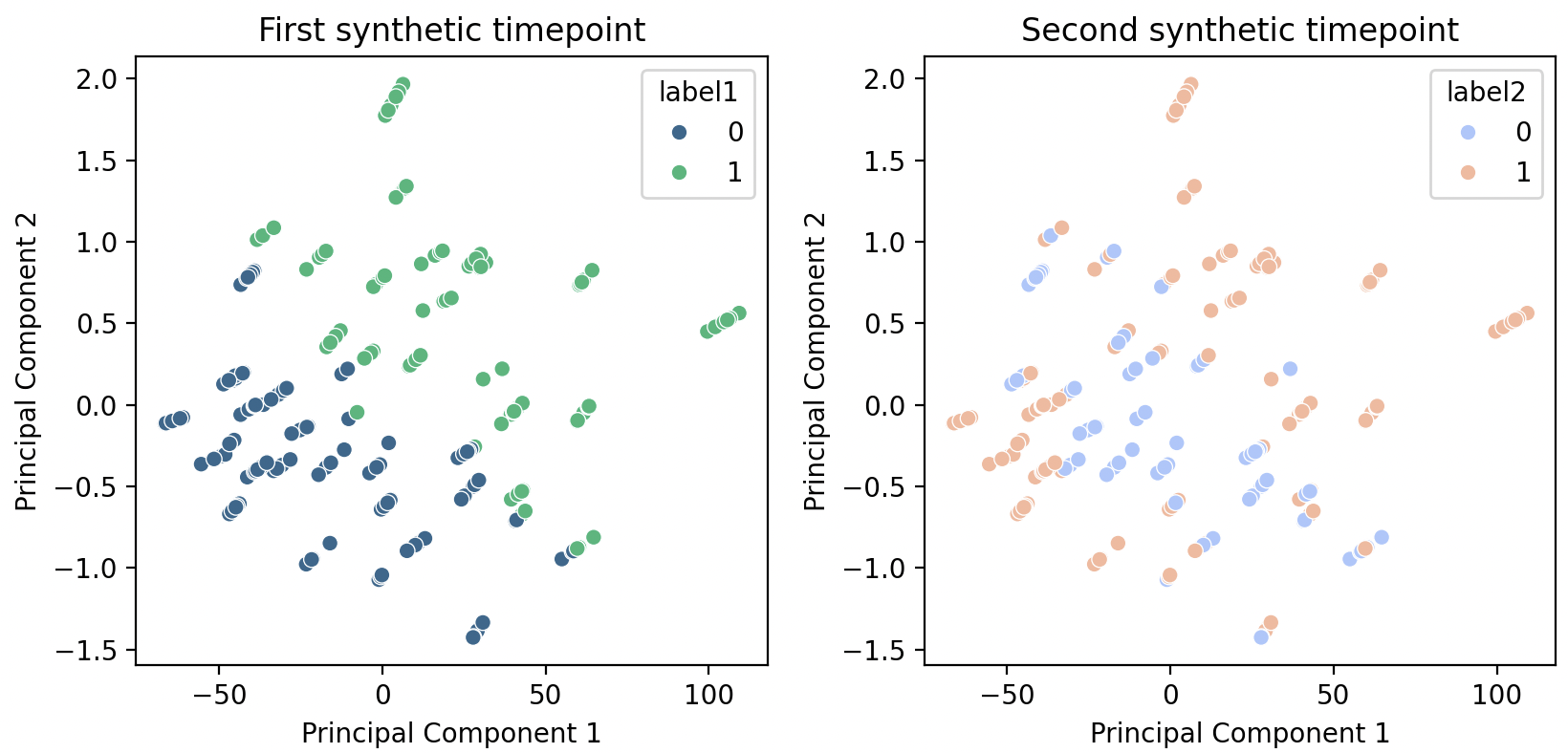}
	\caption[Synthetic longitudinal dataset overview.]{Synthetic longitudinal dataset overview. Left panel: PCA representation of the synthetic dataset colored by the target label at first timepoint. Right panel: PCA representation of the synthetic dataset colored by the target label at second timepoint.}
	\label{fig:synthetic_data_pca}
\end{figure}

To simulate the second follow-up response, we introduced transition probabilities inspired by the Lumiere dataset, but we deliberately emphasized them to obtain a clearer stratification of patients and a more homogeneous dataset, accentuating transitions which are underrepresented in the original data, such as the shift from progressive to stable disease. Specifically, we identified four score ranges, $(-\infty, -1)$, $[-1, 0)$, $[0, 1]$, and $(1, +\infty)$, obtained for the first synthetic follow-up response.
All cases with extreme positive scores, indicating progression, remain progressive. The majority of patients who were stable at the first follow-up, i.e., with extreme negative score at the first follow-up, transition to a progressive state. These align with the trends observed in the Lumiere dataset. Additionally, we introduced the possibility that uncertain or borderline progressive cases (i.e., those with low positive progression scores at the first follow-up) may revert to a stable condition. 
Therefore, the labeling for the second follow-up response was conducted as follows.
\begin{itemize}
	\item All the patients having scores in $(1, +\infty)$ were labeled as $1$, reflecting the persistence of a severe condition (i.e., progressive patients remaining progressive).
	\item $90$\% of patients were sampled from the extreme low scores $(-\infty, -1)$. These patients were labeled as $1$, reflecting a transition of patient condition from the first to the second follow-up (i.e., stable patients becoming progressive).
	\item All patients having scores in $[-1, 0]$ were labeled as $0$, simulating stabilization of patient condition between first and second follow-up (i.e., stable patients remain stable).
	\item $50$\% of patients were sampled from the range of moderate positive scores $[0, 1]$ and labeled as $0$, simulating partial improvement in patient condition between first and second follow-up (i.e., transition from progressive to stable disease).
\end{itemize}
Principal Component Analysis (PCA) representation of the synthetic dataset is shown in Figure \ref{fig:synthetic_data_pca}.

\section{Benchmark experiments}\label{sec:benchmark}
	In order to benchmark the proposed probabilistic model for longitudinal prediction of treatment response, we conducted a series of experiments to predict the second follow-up response using radiomics from Lumiere dataset and the custom-designed synthetic dataset. Our goal was to assess whether our longitudinal approach could outperform existing radiomics-based methods for longitudinal prediction.
	Specifically, we trained the following models to predict the second follow-up response to treatment (model names derive from the set of features used for training). With the Lumiere dataset:
	\begin{itemize}
			\item prediction of the second follow-up response using baseline radiomic features, 'baseline radiomics' model (also referred to as $f_{2}^i$, in Section \ref{sec:longitudinal_approach});
			\item prediction of the second follow-up response using the true response labels at the first follow-up (i.e., RANO labels), 'true labels at $1$st follow-up' model;
			\item prediction of the second follow-up response using radiomics extracted from the first follow-up imaging, when available, 'radiomics at $1$st follow-up' model;
			\item prediction of the second follow-up response with delta radiomics (relative change between features at first follow-up and baseline), 'delta radiomics' model.
	\end{itemize}
	
	A model based on radiomic feature concatenation at the first and second follow-ups was not considered due to the excessive number of features relative to the limited number of patients. 
	
	Similarly, for the synthetic dataset:
		\begin{itemize}
		\item prediction of the second follow-up response using baseline synthetic features, 'baseline synthetic' model (also referred to as $f_{2}^i$, in Section \ref{sec:longitudinal_approach});
		\item prediction of the second follow-up response using the synthetic response labels at the first follow-up, 'synthetic labels at 1st follow-up' model.
	\end{itemize}
	
	The performances of these models were compared with the ones of the longitudinal model $f_{2}^{L, i}$, described in Section \ref{sec:longitudinal_approach}. We recall that this model was trained with baseline features (radiomic or synthetic features) and the feature 'response at $1$st follow-up'. The testing phase, to evaluate model performance, was performed with various approaches to modeling the feature 'response at $1$st follow-up'.  The longitudinal models were named according to the way this feature was modeled. Specifically, in both the Lumiere and synthetic case, the feature 'response at $1$st follow-up' was modeled:
	\begin{itemize}
		\item by using the true response labels at the first follow-up, $y_{1, true}^{i,j}$, leading to 'baseline + true labels' and 'baseline + synthetic labels' models, using Lumiere and synthetic dataset, respectively;
		\item by using the predicted response labels at the first follow-up, $y_1^{i,j}$, obtaining the 'baseline + predicted labels' model (we use the same model name for both Lumiere and synthetic case);
		\item by using the sampled response label $\hat{y}_1^{j}$, obtained as described in Section \ref{sec:longitudinal_approach}, producing the 'baseline + G-KDE labels' model, that is our proposed method (we use the same model name for both Lumiere and synthetic case). 
	\end{itemize}
	
To compute the predicted response labels at the first follow-up, $y_1^{i,j}$ and the sampled response label $\hat{y}_1^{j}$, we trained a model for the prediction of the first follow-up response with baseline features, referred to as 'baseline radiomics' or 'baseline synthetic' model, depending on the dataset used for training (either Lumiere or synthetic), also indicated with $f_{1}^i$ in Section \ref{sec:longitudinal_approach}. The performance of this model is reported among the results; however, it should not be directly compared to the longitudinal prediction at the second follow-up. Instead, it may serve as a reference point for understanding the difference in the predictive power of baseline features for forecasting the response to treatment at the first versus the second follow-up.

All the models in this study that were trained with Lumiere radiomics included additional demographics features, age at surgery and sex, along with the feature 'number of months to the first (second) post-op follow-up', to account for variations in follow-up intervals across patients when predicting the first (second) post-operative treatment response.
A summary of the models employed for this study is reported in Table \ref{tab:performed_experiments}.

	{\fontsize{9}{9}\selectfont
	\begin{longtable}{|l|c|c|c|}
		\caption[Summary of the performed experiments]{Summary of the performed experiments. The 'Model' column indicates the model names. The 'Longitudinality' column specifies whether the model leverages information from multiple timepoints, indicating a truly longitudinal approach. The 'Synthetic' and 'Lumiere' columns indicate whether the models were trained on the respective datasets.}
		\label{tab:performed_experiments} \\
		
		\hline
		\textbf{Model} & \textbf{Longitudinality} & \textbf{Synthetic} & \textbf{Lumiere} \\
		\hline
		\endfirsthead
		
		\hline
		\textbf{Model} & \textbf{Longitudinality} & \textbf{Synthetic} & \textbf{Lumiere} \\
		\hline
		\endhead
		
		\hline \multicolumn{4}{|r|}{\textit{Table \ref{tab:performed_experiments} continued on next page}} \\
		\hline
		\endfoot
		
		\hline
		\endlastfoot
		
		Baseline     $\{$radiomics, synthetic$\} $                   & \ding{55} & \ding{51} & \ding{51}   \\ \hline		 $\{$True, synthetic$\}$ labels at $1$st follow-up          &      \ding{55}      & \ding{51} & \ding{51}  \\ \hline
		Radiomics at $1$st follow-up          & \ding{55} &     \ding{55}      & \ding{51}   \\ \hline
		Delta radiomics    & \ding{51} & \ding{55} &   \ding{51}        \\ \hline
		Baseline + $\{$true, synthetic$\}$ labels              & \ding{51} & \ding{51} & \ding{51}  \\ \hline
		Baseline + predicted labels              &    \ding{51}       & \ding{51} & \ding{51} \\ \hline
		Baseline + G-KDE labels    & \ding{51} & \ding{51} &   \ding{51}        \\ \hline
		
	\end{longtable}
	}

\clearpage
	\section{Machine learning models}\label{sec:ml_models}

	The dataset was initially divided into stratified $60$\%-$40$\% splits for training and testing, with stratification based on the responses to treatment at the first and second follow-ups ($40$ splits for Lumiere dataset, $230$ splits for the synthetic dataset).
	
	For Lumiere dataset, the imbalanced data issue was addressed via inverse frequency weighting-class strategy for prediction of the first follow-up response, when the dataset is more balanced ($32$ SD, $15$ PD). SMOTE was used for the prediction of the second follow-up response ($11$ SD and $36$ PD when using baseline radiomics, $6$ SD and $33$ PD when using radiomics from $1$st follow-up imaging). For the synthetic dataset, inverse frequency weighting-class strategy is used for both predictions, for first and second follow-up response. 
	Features were normalized between $0$ and $1$ before training.
	
	All classifiers were implemented as Logistic Regression models with L$1$-norm penalty and the regularization parameter $C$ was set to $1$. Leave-one-out cross-validation was used.  
	Baseline and longitudinal prediction models were evaluated via the following performance metrics: accuracy, recall, specificity, balanced accuracy, and ROC-AUC. 
	
	To check if that the model $f_{1}^i$ could generate reliable probability estimates for the response at the first follow-up, we evaluated the calibration of this model. A well-calibrated classifier should produce probability estimates that can be directly interpreted as confidence levels, meaning that if the model predicts a probability of $0.8$, about $80$\% of those cases should be positive. In this study, model calibration is useful to consider, as the construction of the proposed longitudinal model, $f_{2}^{L, i}$, relies on the probabilities associated with the predicted response at the first follow-up, generated by the model $f_{1}^i$. Log loss \cite{bishop2006pattern} and Brier score \cite{brier} are two widely-used metrics for evaluating the calibration of probabilistic models. Log loss, also known as cross-entropy loss, is defined as
	$$ \text{LogLoss} = -\frac{1}{N} \sum_{i=1}^{N} \left( y_i \log(p_i) + (1 - y_i) \log(1 - p_i) \right),$$
	where \( y_i \) is the true label ($0$ or $1$) and \( p_i \) is the predicted probability for sample \( i \). Log loss penalizes confident but wrong predictions very heavily, and rewards confident correct predictions, making it particularly sensitive to miscalibrations in high-confidence regions. The Brier score is defined as
	$$\text{Brier score} = \frac{1}{N} \sum_{i=1}^{N} (p_i - y_i)^2,$$
	and measures the mean squared difference between predicted probabilities and the actual outcomes. It treats overconfident and underconfident errors more symmetrically than log loss, and is generally less sensitive to extreme mistakes. 
	The calibration of the baseline model $f_{1}^i$ was assessed using the Brier score and the Log Loss, both on the original model and after isotonic calibration, for comparison purposes.
	Calibrating a classifier consists of fitting a regressor (calibrator) that maps the output of the classifier to a calibrated probability. The calibrator tries to predict the conditional event probability $\mathbb{P}(y_i=1|p_i)$. We employed the isotonic regression \cite{isotonic}, a non-parametric method, to calibrate our model $f_{1}^i$. 
	In isotonic calibration, we seek a function \( g \) that maps the uncalibrated predicted probabilities \( p_i \) to calibrated probabilities \( \hat{p}_i \), such that $\hat{p}_i = g(p_i)$, where $g$ is a monotonic non-decreasing mapping. The function $g$ is found by minimizing the squared error between adjusted probabilities and actual outcomes:
	$$\sum_{i=1}^{N} \left( g(p_i) - y_i \right)^2.$$
	Isotonic regression captures complex, non-smooth calibration patterns, 
	and is suited for correcting arbitrary calibration errors. Instead, for smooth and global miscalibration, Platt's scaling calibration \cite{platt} would be more appropriate. The models were developed in Python (v$3$.$12$.$2$).

\section{Numerical results and discussion}\label{sec:results_longitudinal}

The eligible subjects from Lumiere dataset present variable follow-up intervals between subsequent post-operative timepoints (mean interval between the $1$st and $2$nd post-operative follow-up $108.7$ days, std $66.4$ days, min $21$ days, max $343$ days). At the first follow-up, there are $32$ SD and $15$ PD responses, while at the second follow-up $11$ SD and $36$ PD cases are registered. Specifically, $21$ patients transition from SD to PD, $11$ remain SD, $0$ transition from PD to SD, and $15$ PD cases remain PD.

The results presented on Lumiere dataset are obtained from pre-contrast T$1$-weighted radiomics, as models based on FLAIR, T$2$, and post-contrast T$1$ yield random baseline predictions at the first follow-up, therefore are not suitable for meaningful analysis or longitudinal assessments.

The estimation of the probability distribution of response 1 (i.e., the probability distribution of progressive disease for Lumiere dataset) relies on the assumption that the probabilities generated by the baseline model to predict the response at the first follow-up can be directly interpreted as event likelihoods, i.e., a predicted value of $0.7$ truly reflects a $70$\% probability of the event occurring. For such reason we evaluated model calibration via Brier score and Log Loss.

\begin{figure}[ht]
		\centering
		\includegraphics[width=0.4\linewidth]{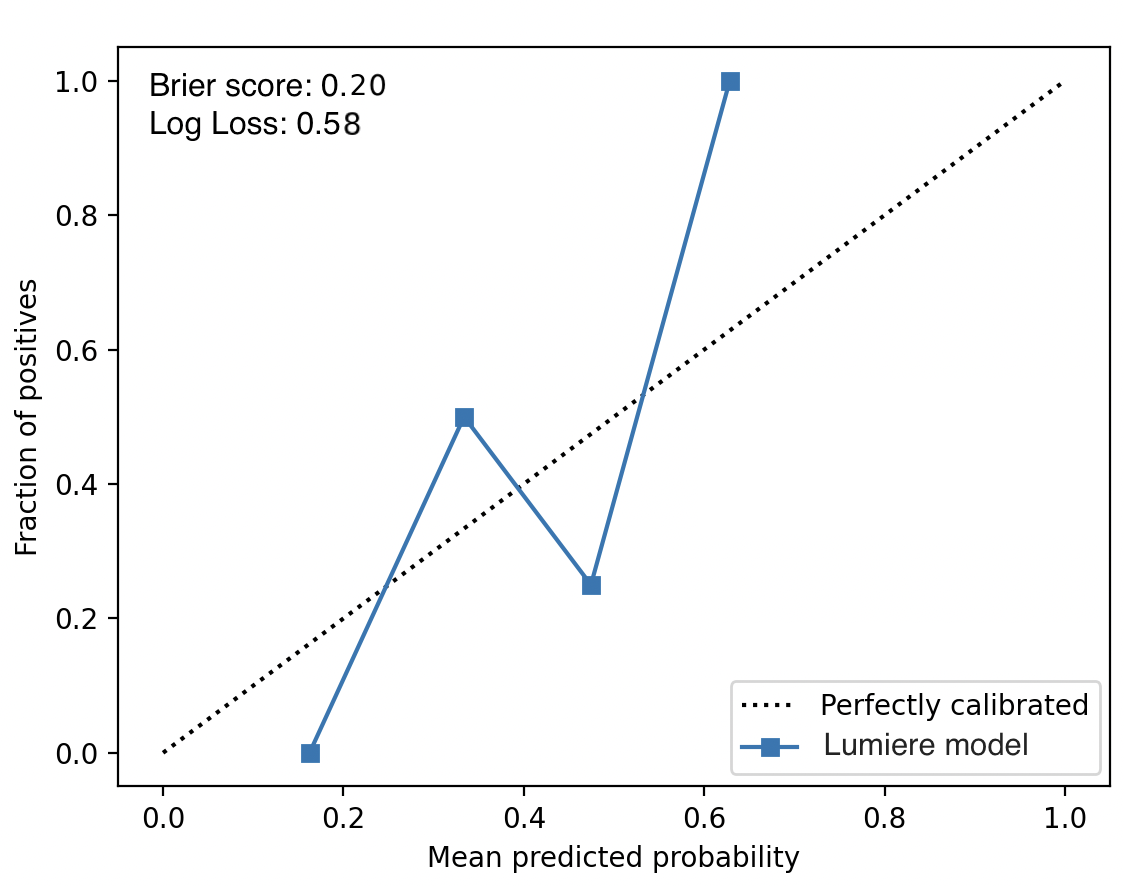}\quad
		\includegraphics[width=0.4\linewidth]{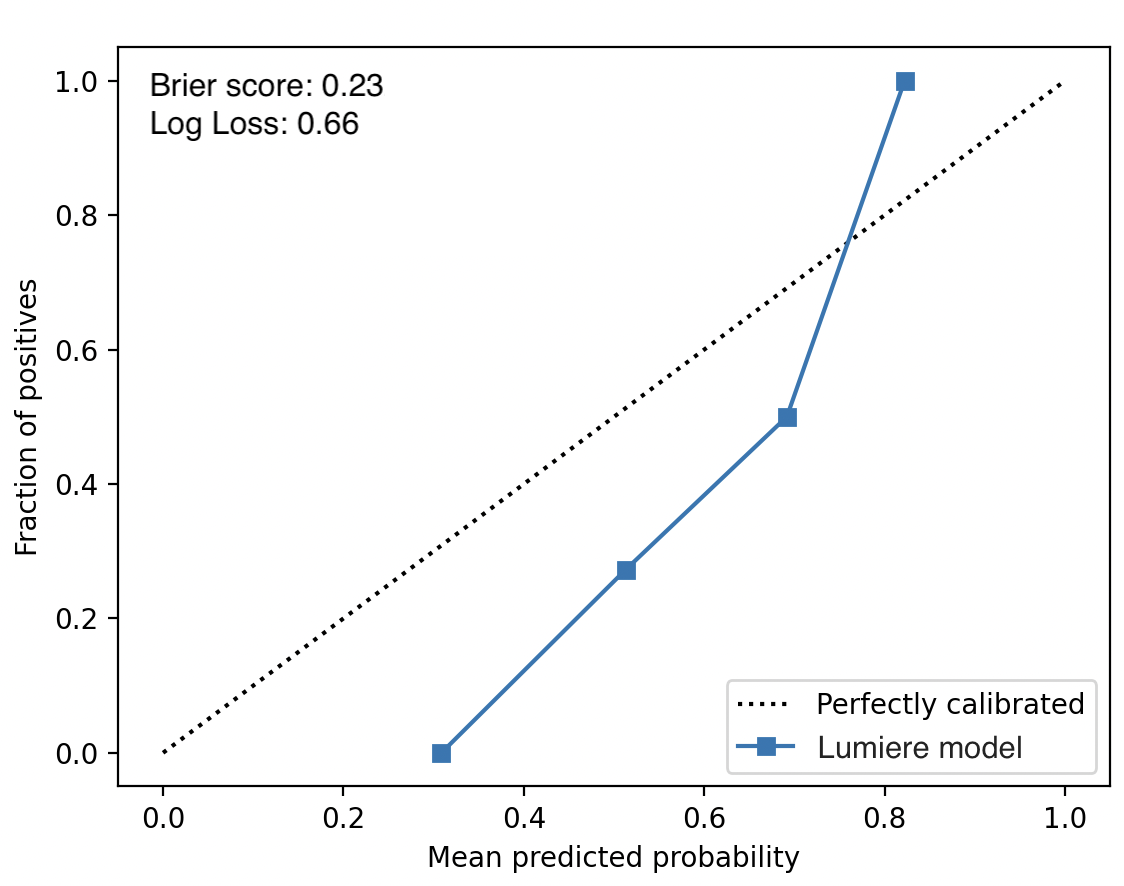}\\
		\includegraphics[width=0.4\linewidth]{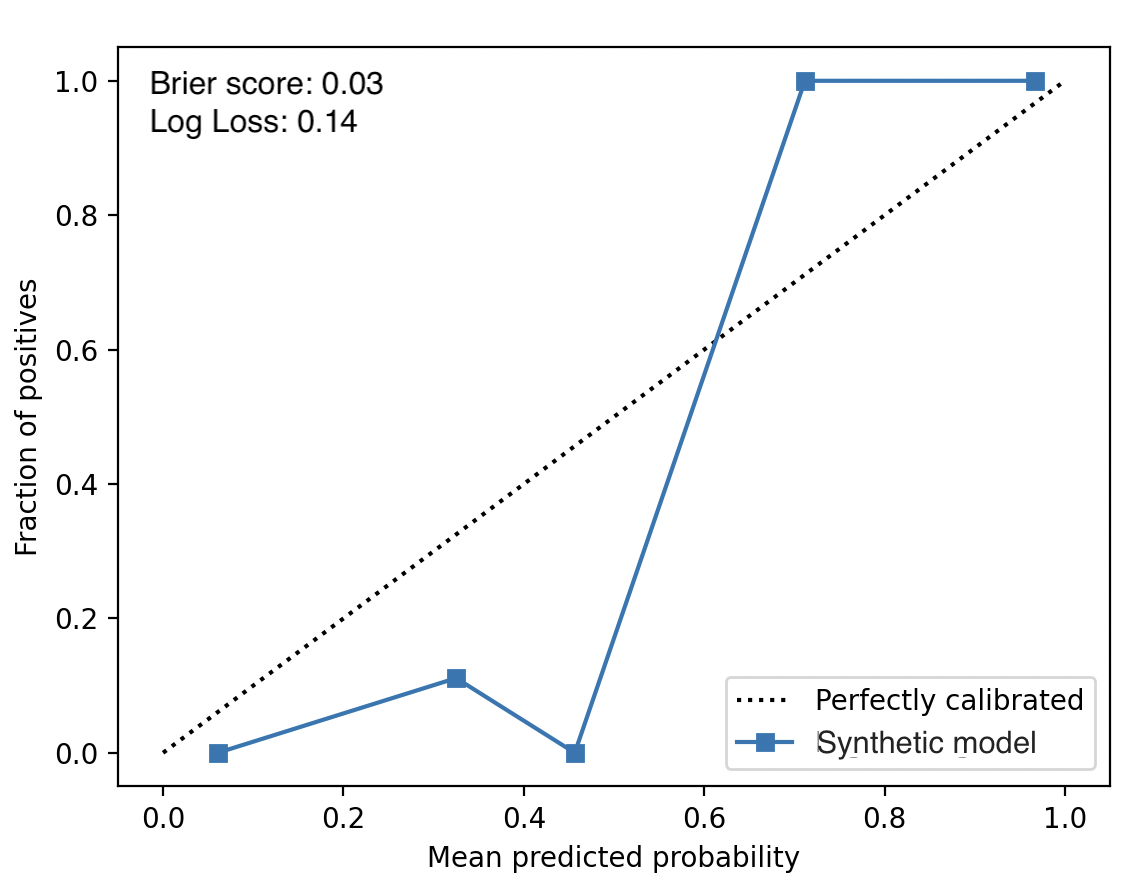}\quad
		\includegraphics[width=0.4\linewidth]{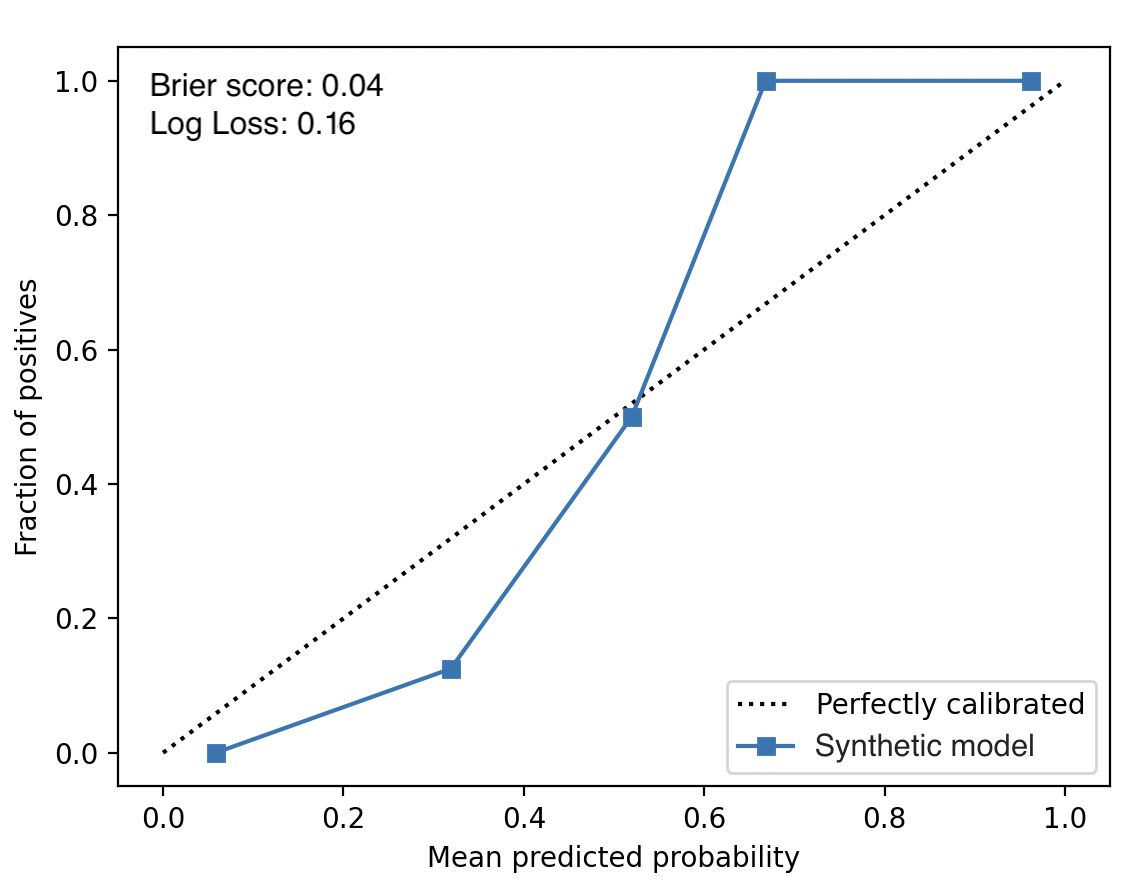}\\
	
		\caption[Calibration curves for the baseline model to predict the response at first follow-up, in the case of synthetic and Lumiere dataset.]{Calibration curves for the baseline model to predict the response at first follow-up. The curves are generated by grouping predicted probabilities into bins and then plotting the average predicted probability for each bin against the corresponding observed frequency (fraction of positive cases). Top row: calibration curve of two model trained on Lumiere dataset. Bottom row: calibration curves of two models trained on the synthetic dataset.}
		\label{fig:calibration}
\end{figure}

Table \ref{tab:baseline_model_calibration} presents the Brier scores and Log Loss for the baseline models to predict the response at the first follow-up, $f_1^i$, trained on synthetic and Lumiere datasets. For comparison, the same metrics are provided for the corresponding models after isotonic calibration. 
Table \ref{tab:baseline_model_calibration} shows that both scores resulted lower for models trained on the synthetic dataset than for Lumiere-based models, indicating a better overall calibration in the synthetic setting. This could happen because the Lumiere training size is too small, overfitting case, or because the model is misspecified, e.g., if the true decision boundary of the dataset is a highly non-linear function of the input features. For comparison, the same metrics are provided for the corresponding models after isotonic calibration. Although applying isotonic calibration to the models slightly lowered the mean Brier score for both dataset, the Log Loss resulted to increase significantly. This deteriorates model predictive effectiveness, as the logistic regression classifiers are trained by minimizing the Log Loss, undermining the purpose of calibration itself, and is likely due to data scarcity, as the calibration dataset (in this case, the test set) does not properly match the real distribution of the data. Figure \ref{fig:calibration} shows examples of calibration curves for the baseline model, to predict the response at first follow-up, having under-confident trend, for both synthetic and Lumiere dataset.

\renewcommand{\arraystretch}{1.5}
{\fontsize{10}{10}\selectfont
	\begin{longtable}[b]{|lcc|}
		\caption[Calibration scores for baseline models to predict the response at the first follow-up, in the case of synthetic and Lumiere dataset.]{Calibration scores for baseline models to predict the response at the first follow-up, before and after isotonic calibration, in the case of synthetic and Lumiere dataset. Mean and standard deviations are computed across all splits.}
		\label{tab:baseline_model_calibration} \\
		\hline
		\textbf{Model}	& \textbf{Brier Score}                    & \textbf{Log Loss }      \\ \hline
		\endfirsthead
		\hline
		\textbf{Model}	& \textbf{Brier Score}                    & \textbf{Log Loss }     \\ \hline
		\endhead
		\hline \multicolumn{3}{r}{\textit{Table \ref{tab:baseline_model_calibration} continued on next page}} \\ 
		\endfoot
		\hline
		\endlastfoot
		Baseline synhetic             & $0.05 \pm 0.01$ & $0.18 \pm 0.02$ \\ \hline
		Baseline synthetic + isotonic calibration & $0.03 \pm 0.02$ & $0.41 \pm 0.48$ \\ \hline
		Baseline radiomics             & $0.22 \pm 0.02$ & $0.64 \pm 0.05$ \\ \hline
		Baseline radiomics + isotonic calibration   & $0.22 \pm 0.06$ & $4.13 \pm 2.70$ \\ \hline
	\end{longtable}
}

\begin{figure}[h!]
	\centering
	\includegraphics[width=0.45\linewidth]{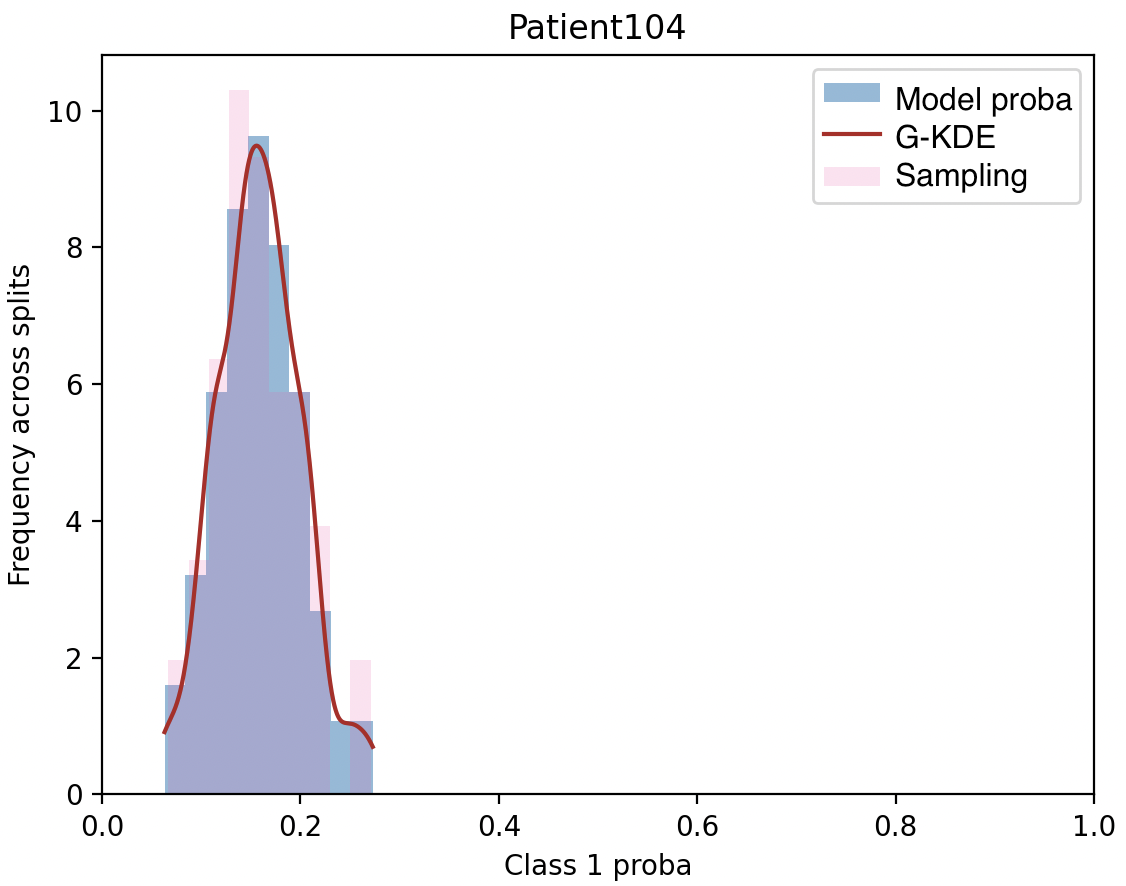}\quad
	\includegraphics[width=0.43\linewidth]{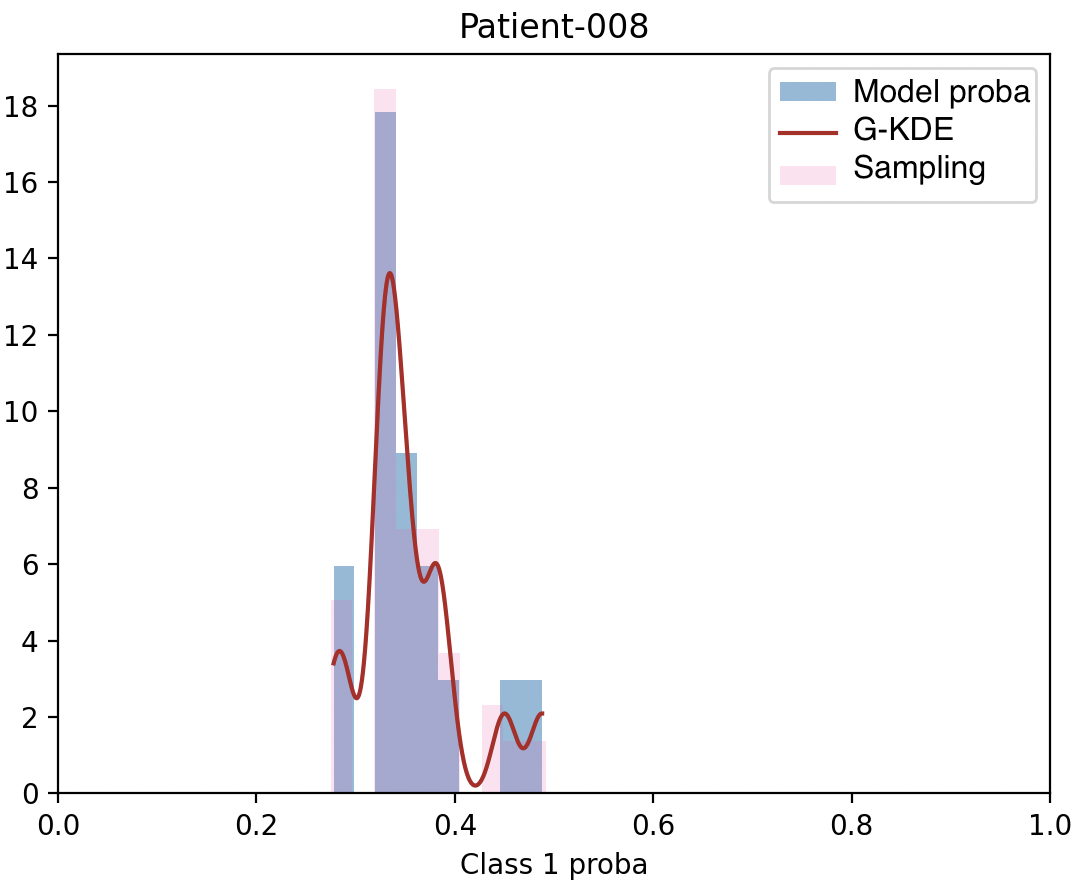}\\
	\includegraphics[width=0.45\linewidth]{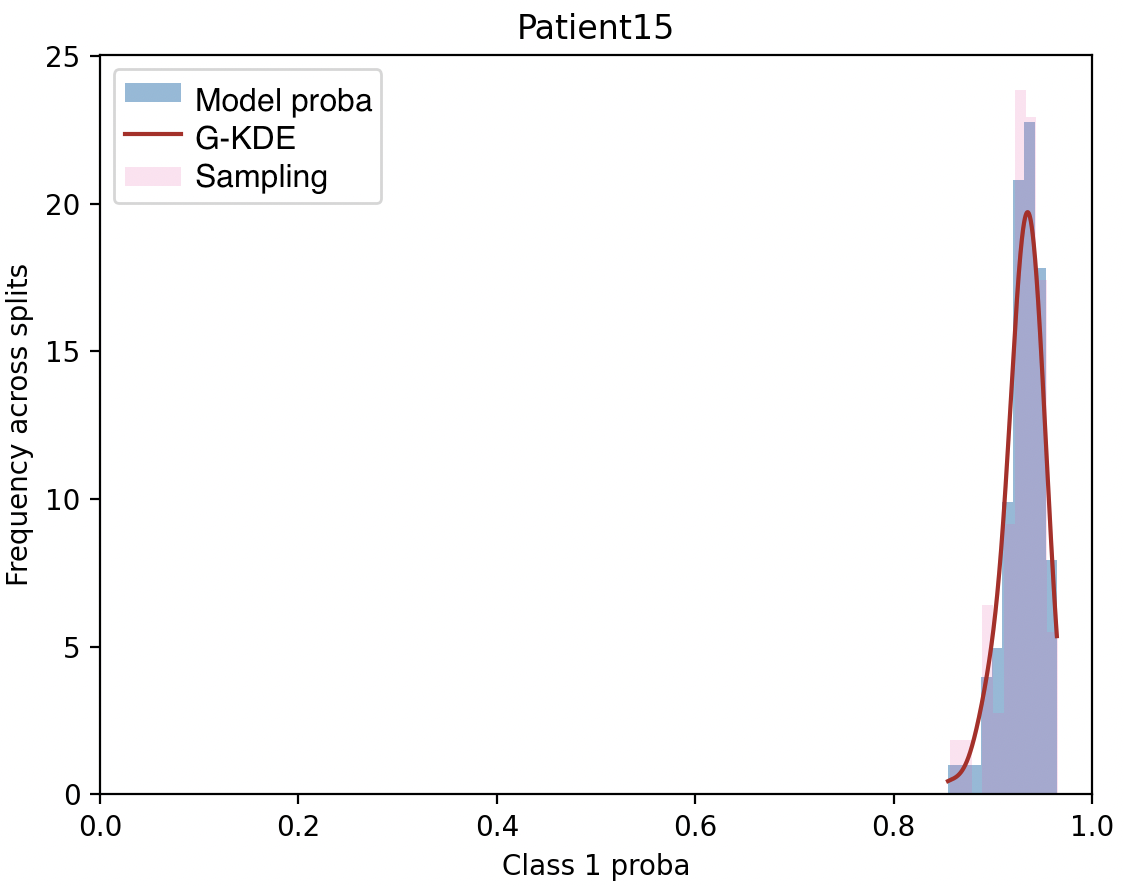}\quad
	\includegraphics[width=0.43\linewidth]{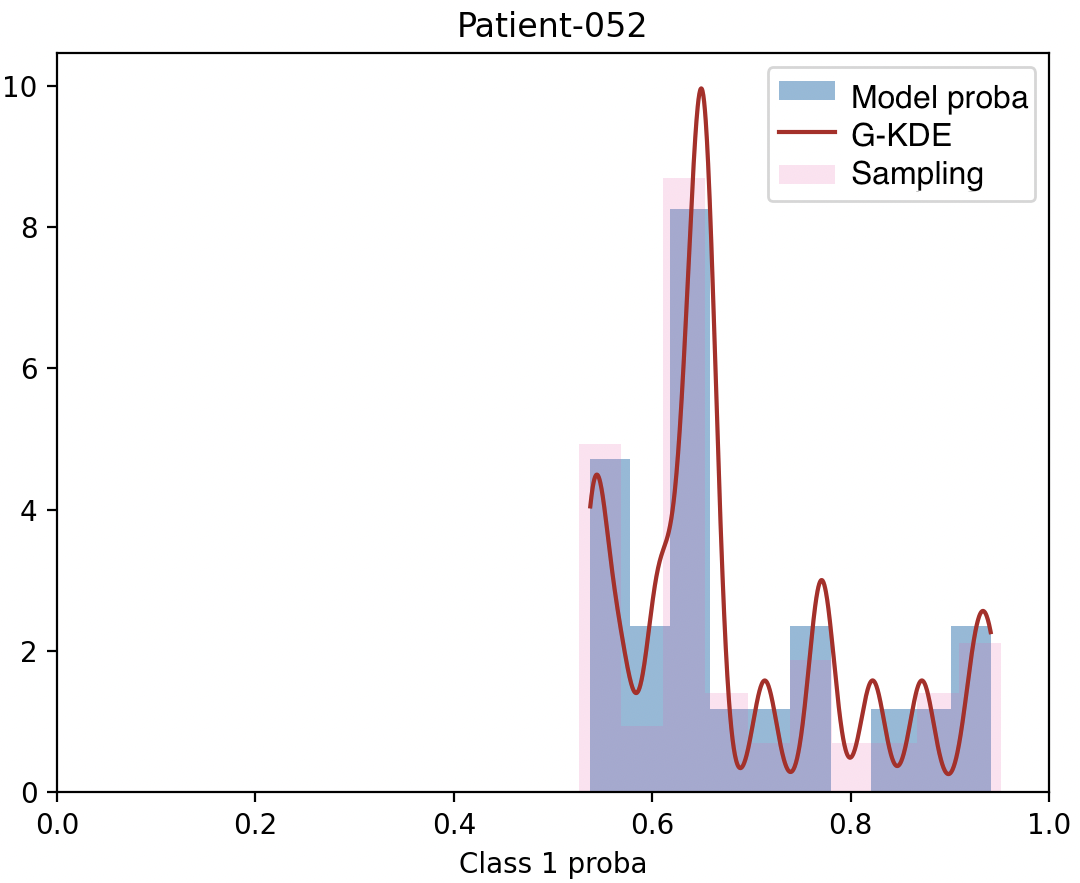}\\
	\includegraphics[width=0.45\linewidth]{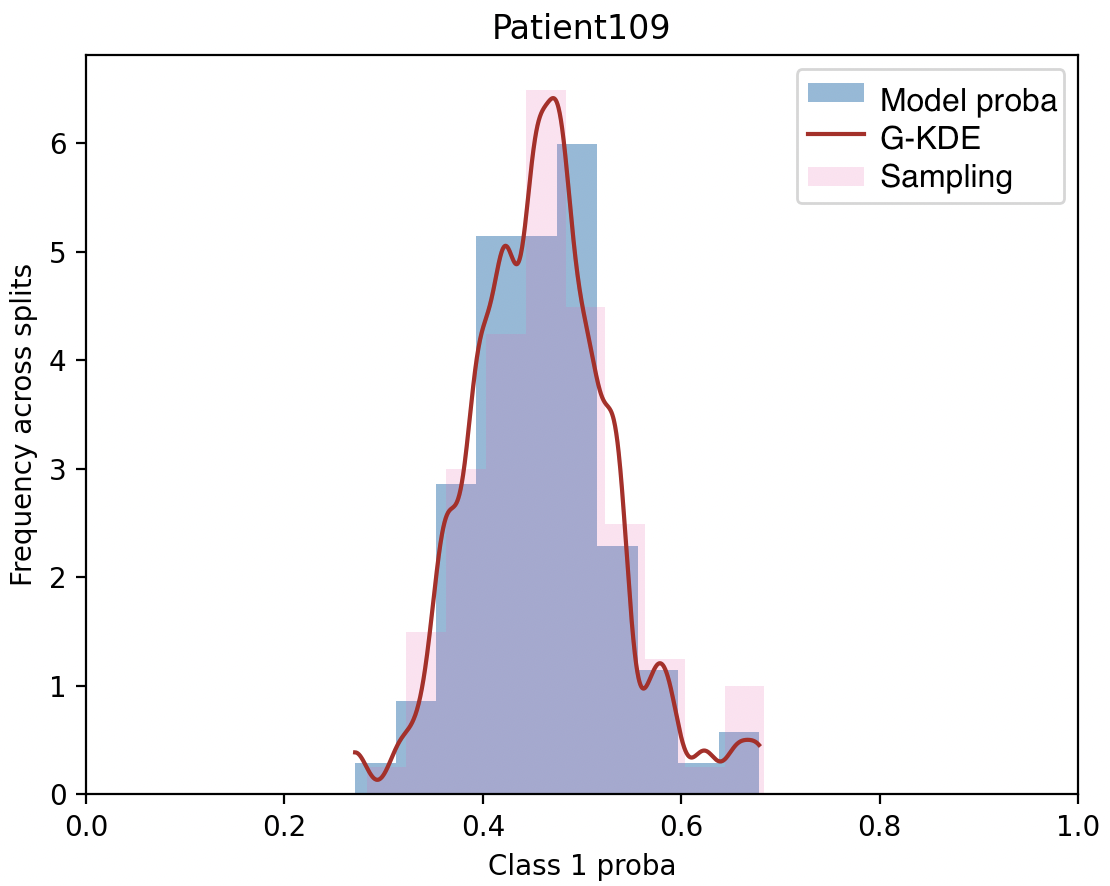}\quad
	\includegraphics[width=0.43\linewidth]{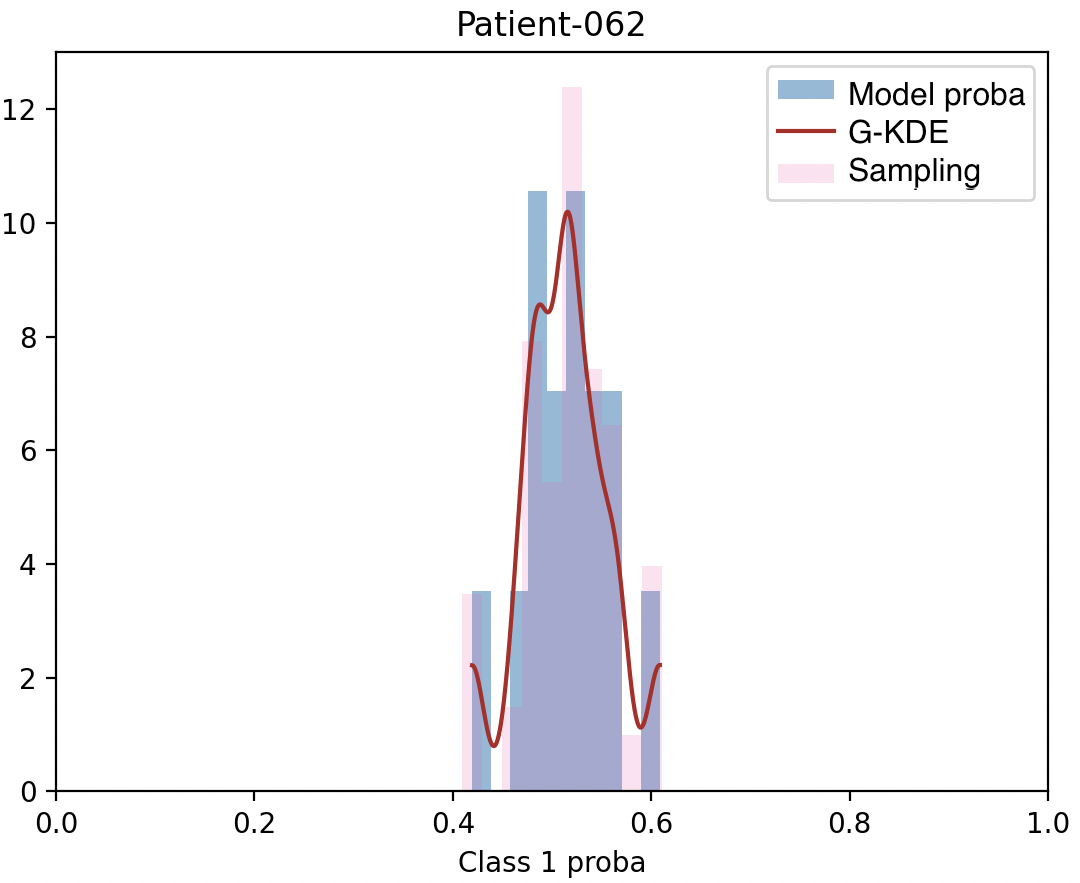}\\
    \caption[Examples of probability distribution of disease progression at the first follow-up, for patients from synthetic and Lumiere dataset.]{Examples of probability distribution of disease progression (response of class $1$) at the first follow-up. Light blue: normalized histogram (area $1$) of $\{\texttt{probas}_1^{i,j}\}_i$ obtained across splits, $i$, for each patient, $j$. Red: probability density function (PDF) of PD distribution estimated via G-KDE, $\hat{Y}^j_1$, for the $j$-th patient. Pink: samples drawn from the G-KDE PDF, $\{\widehat{\texttt{proba}}_1^{j}\}$. First column: examples from the synthetic dataset. Second column: examples from Lumiere dataset.}
	\label{fig:probas}
\end{figure}

Figure \ref{fig:probas} depicts the probability distribution of response=1 at the first follow-up for various patients, provided by the baseline models $f_1^i$, in both the synthetic and Lumiere dataset case. This corresponds to the probability of disease progression (PD) for Lumiere dataset. The figure also shows examples of the fit provided by G-KDE, $\hat{Y}^j_1$, along with data samples drawn from $\hat{Y}^j_1$. 

Tables \ref{tab:skill_scores_longitudinal_synthetic} and \ref{tab:skill_scores_longitudinal_T1} report the mean skill-scores and their $95$\% confidence interval obtained by testing each ML model (baseline, longitudinal, and benchmark models), on the synthetic and Lumiere dataset, respectively. A summary of the implemented models is in Table \ref{tab:performed_experiments}. 
We highlight that the first row of both table describes the prediction of response at the first follow-up, which is then leveraged for the probabilistic modeling of the longitudinal response at the second follow-up. 
SMOTE augmentation did not have a noticeable impact on the predictive performance with the synthetic dataset and for predicting response at first follow-up using Lumiere dataset. 
In contrast, for the prediction of the second follow-up response with the Lumiere dataset, inverse frequency class weighting alone was insufficient to address class imbalance, resulting in performance close to random. The application of SMOTE in this case led to the results reported in Table \ref{tab:skill_scores_longitudinal_T1}.

\setlength{\tabcolsep}{1pt} 
\renewcommand{\arraystretch}{2.5}
 {\fontsize{8}{8}\selectfont
	\begin{longtable}{|lccccc|}
		\caption[Performance of the implemented models on the synthetic dataset.]{Performance of the implemented models on the synthetic dataset. Mean skill-scores and their $95$\% confidence interval (within brackets) are reported. Longitudinal models are differentiated by the modeling of 'response at $1$st follow-up' feature: 'baseline + synthetic labels' for the true synthetic responses $y_{1, true}^{i,j}$, 'baseline + predicted labels' for predicted responses $y_1^{i,j}$, and 'baseline + G-KDE labels' for sampled responses $\hat{y}_1^{j}$. The performance of proposed longitudinal approach ('baseline + G-KDE labels') is reported in bold.}
		\label{tab:skill_scores_longitudinal_synthetic} \\
		\hline
		\textbf{Model}             & \textbf{Accuracy}    & \makecell{\textbf{Balanced} \\\textbf{accuracy}} & \textbf{Recall}      & \textbf{Specificity} & \textbf{ROC-AUC}     \\ \hline
		\endfirsthead
		\hline
		\textbf{Model}             & \textbf{Accuracy}    & \makecell{\textbf{Balanced} \\ \textbf{accuracy}} & \textbf{Recall}      & \textbf{Specificity} & \textbf{ROC-AUC}     \\ \hline
		\endhead
		\hline \multicolumn{6}{r}{\textit{Table \ref{tab:skill_scores_longitudinal_synthetic} continued on next page}} \\ 
		\endfoot
		\hline
		\endlastfoot

        \multicolumn{6}{|c|}{\textbf{Prediction of $\boldsymbol{1}$st follow-up response}} \\ \hline
        \makecell[l]{Baseline synthetic}  & \makecell{$0.947$ \\ $(0.944, 0.950)$} & \makecell{$0.950$ \\ $(0.947, 0.953)$} & \makecell{$0.967$ \\ $(0.963, 0.971)$} & \makecell{$0.933$ \\ $(0.928, 0.937)$} & \makecell{$0.994$ \\ $(0.993, 0.994)$} \\ \hline
        \multicolumn{6}{|c|}{\textbf{Prediction of $\boldsymbol{2}$nd follow-up response}} \\ \hline
		\multicolumn{6}{|c|}{\textit{Benchmark models}}\\ \hline
        \makecell[l]{Baseline synthetic}  & \makecell{$0.603$ \\ $(0.598, 0.608)$} & \makecell{$0.620$ \\ $(0.614, 0.626)$} & \makecell{$0.540$ \\ $(0.533, 0.548)$} & \makecell{$0.699$ \\ $(0.686, 0.712)$} & \makecell{$0.707$ \\ $(0.703, 0.711)$} \\ \hline
		\makecell[l]{Synthetic labels \\ at $1$st follow-up } & \makecell{$0.628$ \\ $(0.627, 0.628)$} & \makecell{$0.655$ \\ $(0.654, 0.655)$} & \makecell{$0.527$ \\ $(0.527, 0.528)$} & \makecell{$0.782$ \\ $(0.781, 0.783)$} & \makecell{$0.655$ \\ $(0.654, 0.655)$} \\ \hline
        \multicolumn{6}{|c|}{\textit{Proposed longitudinal models}} \\ \hline
		\makecell[l]{Baseline + \\synthetic labels }  & \makecell{$0.731$ \\ $(0.728, 0.734)$} & \makecell{$0.739$ \\ $(0.736, 0.741)$} & \makecell{$0.704$ \\ $(0.699, 0.709)$} & \makecell{$0.773$ \\ $(0.771, 0.775)$} & \makecell{$0.801$ \\ $(0.798, 0.803)$} \\ \hline
		\makecell[l]{Baseline + \\predicted labels}  & \makecell{$0.697$ \\ $(0.694, 0.700)$} & \makecell{$0.696$ \\ $(0.693, 0.700)$} & \makecell{$0.699$ \\ $(0.694, 0.704)$} & \makecell{$0.693$ \\ $(0.689, 0.698)$} & \makecell{$0.744$ \\ $(0.740, 0.748)$} \\ \hline
		\makecell[l]{\textbf{Baseline + }\\ \textbf{G-KDE labels}}  & \makecell{$\boldsymbol{0.696}$ \\ $\boldsymbol{(0.693, 0.699)}$} & \makecell{$\boldsymbol{0.696}$ \\ $\boldsymbol{(0.692, 0.699)}$} & \makecell{$\boldsymbol{0.697}$ \\ $\boldsymbol{(0.692, 0.702)}$} & \makecell{$\boldsymbol{0.694}$ \\ $\boldsymbol{(0.690, 0.699)}$} & \makecell{$\boldsymbol{0.743}$ \\ $\boldsymbol{(0.739, 0.747)}$} \\ \hline

	\end{longtable}
}

\setlength{\tabcolsep}{1pt} 
\renewcommand{\arraystretch}{2.5}
{\fontsize{8}{8}\selectfont
	\begin{longtable}{|lccccc|}
		\caption[Performance of the implemented models on Lumiere dataset.]{Performance of the implemented models on Lumiere dataset. Mean skill-scores and their $95$\% confidence interval (within brackets) are reported. Longitudinal models are differentiated by the modeling of 'response at $1$st follow-up' feature: 'baseline + true labels' for actual treatment responses $y_{1, true}^{i,j}$, 'baseline + predicted labels' for the predicted responses $y_1^{i,j}$, and 'baseline + G-KDE labels' for sampled responses $\hat{y}_1^{j}$. The performance of proposed longitudinal approach ('baseline + G-KDE labels') is reported in bold.}
		\label{tab:skill_scores_longitudinal_T1} \\
		\hline
		\textbf{Model}             & \textbf{Accuracy}    & \makecell{\textbf{Balanced} \\ \textbf{accuracy}} & \textbf{Recall} & \textbf{Specificity}      & \textbf{ROC-AUC}     \\ \hline
		\endfirsthead
		\hline
		\textbf{Model}             & \textbf{Accuracy}    & \makecell{\textbf{Balanced} \\ \textbf{accuracy}} & \textbf{Recall} & \textbf{Specificity}      & \textbf{ROC-AUC}     \\ \hline
		\endhead
		\hline \multicolumn{6}{r}{\textit{Table \ref{tab:skill_scores_longitudinal_T1} continued on next page}} \\ 
		\endfoot
		\hline
		\endlastfoot

        \multicolumn{6}{|c|}{\textbf{Prediction of $\boldsymbol{1}$st follow-up response}} \\ \hline
        \makecell[l]{Baseline radiomics} & \makecell{$0.721$ \\ $(0.693, 0.749)$} & \makecell{$0.682$ \\ $(0.660, 0.704)$} & \makecell{$0.575$ \\ $(0.524, 0.626)$} & \makecell{$0.788$ \\ $(0.736, 0.841)$} & \makecell{$0.721$ \\ $(0.697, 0.746)$} \\ \hline
        \multicolumn{6}{|c|}{\textbf{Prediction of $\boldsymbol{2}$nd follow-up response}} \\ \hline
        \multicolumn{6}{|c|}{\textit{Benchmark models}} \\ \hline
        \makecell[l]{Baseline radiomics} & \makecell{$0.680$ \\ $(0.653, 0.708)$} & \makecell{$0.561$ \\ $(0.524, 0.599)$} & \makecell{$0.767$ \\ $(0.735, 0.798)$} & \makecell{$0.356$ \\ $(0.282, 0.430)$} & \makecell{$0.591$ \\ $(0.547, 0.634)$} \\ \hline
        \makecell[l]{True labels \\ at $1$st follow-up} & \makecell{$0.662$ \\ $(0.637, 0.687)$} & \makecell{$0.620$ \\ $(0.606, 0.634)$} & \makecell{$0.676$ \\ $(0.639, 0.713)$} & \makecell{$0.564$ \\ $(0.499, 0.629)$} & \makecell{$0.639$ \\ $(0.625, 0.652)$} \\ \hline
        \makecell[l]{Radiomics \\ at $1$st follow-up} & \makecell{$0.742$ \\ $(0.731, 0.754)$} & \makecell{$0.493$ \\ $(0.478, 0.508)$} & \makecell{$0.825$ \\ $(0.811, 0.839)$} & \makecell{$0.161$ \\ $(0.130, 0.191)$} & \makecell{$0.451$ \\ $(0.425, 0.478)$} \\ \hline
        \makecell[l]{Delta radiomics} & \makecell{$0.755$ \\ $(0.742, 0.768)$} & \makecell{$0.526$ \\ $(0.506, 0.546)$} & \makecell{$0.832$ \\ $(0.818, 0.846)$} & \makecell{$0.219$ \\ $(0.181, 0.257)$} & \makecell{$0.601$ \\ $(0.572, 0.631)$} \\ \hline
        \multicolumn{6}{|c|}{\textit{Proposed longitudinal models}}\\ \hline
        \makecell[l]{Baseline + \\true labels} & \makecell{$0.707$ \\ $(0.677, 0.736)$} & \makecell{$0.626$ \\ $(0.587, 0.666)$} & \makecell{$0.765$ \\ $(0.727, 0.803)$} & \makecell{$0.487$ \\ $(0.405, 0.570)$} & \makecell{$0.688$ \\ $(0.646, 0.729)$} \\ \hline
        \makecell[l]{Baseline + \\predicted labels} & \makecell{$0.678$ \\ $(0.650, 0.705)$} & \makecell{$0.590$ \\ $(0.555, 0.624)$} & \makecell{$0.742$ \\ $(0.708, 0.776)$} & \makecell{$0.438$ \\ $(0.370, 0.505)$} & \makecell{$0.635$ \\ $(0.598, 0.672)$} \\ \hline
        \makecell[l]{\textbf{Baseline +} \\ \textbf{G-KDE labels}} & \makecell{$\boldsymbol{0.689}$ \\ $\boldsymbol{(0.663, 0.715)}$} & \makecell{$\boldsymbol{0.609}$ \\ $\boldsymbol{(0.575, 0.643)}$} & \makecell{$\boldsymbol{0.747}$ \\ $\boldsymbol{(0.715, 0.780)}$} & \makecell{$\boldsymbol{0.470}$ \\ $\boldsymbol{(0.401, 0.540)}$} & \makecell{$\boldsymbol{0.639}$ \\ $\boldsymbol{(0.599, 0.679)}$} \\ \hline

	\end{longtable}
}

Baseline models $f_1^i$ for the prediction of the response at the first follow-up exhibit the highest overall performance across the trained prediction models (see first row of Tables \ref{tab:skill_scores_longitudinal_synthetic} and \ref{tab:skill_scores_longitudinal_T1}). The highest performance is achieved on the synthetic dataset, where the models produce more extreme probability outputs (Figure \ref{fig:probas}, first column), while baseline radiomics maintains good predictive power at the first follow-up for the Lumiere dataset, although with larger uncertainty (Figure \ref{fig:probas}, second column). We recall that the probabilities outputted by these models are then used to build the distribution $\hat{Y}_1^{j}$, described in Section \ref{sec:longitudinal_approach}, that is the core of our probabilistic method for logitudinal prediction of treatment response. 
Of note, the performance of the baseline models for the prediction at the first follow-up should not be directly compared to the prediction at the second follow-up. Instead, it can serve as a reference for comparing the predictive power of pure baseline radiomics in forecasting treatment response at the first and second follow-ups. Examining the benchmark models based on baseline radiomics reveals that baseline features are less predictive of treatment response at the second follow-up compared to the first (mean ROC-AUC on the synthetic dataset: $0.707$ vs. $0.994$; mean ROC-AUC on the Lumiere dataset: $0.591$ vs. $0.721$).

In the synthetic setting, the proposed longitudinal model 'baseline + G-KDE labels' performs similar to the 'baseline + predicted labels' model, using the predicted labels at the first follow-up, but slightly worse than the one using the true synthetic labels, 'baseline + synthetic labels' model. This is expected, since the true synthetic labels do not account for uncertainty in the response at the first follow-up. G-KDE also performs better than the model trained with the $1$st follow-up labels alone, 'synthetic labels at $1$st follow-up' model, indicating that incorporating baseline radiomics in the model provides valuable information to the longitudinal outcome ($0.655$ vs. $0.696$ mean balanced accuracy, $0.655$ vs. $0.743$ mean ROC-AUC). Additionally, G-KDE outperforms the 'baseline synthetic' model, which uses pure baseline synthetic features to forecast the response at the second follow-up ($0.620$ vs. $0.696$ mean balanced accuracy, $0.707$ vs. $0.743$ mean ROC-AUC).

As for Lumiere dataset, 'baseline + G-KDE labels' performance results lower compared to the longitudinal model based on true labels, 'baseline + true labels', but demonstrates superior performance compared to 'baseline + predicted labels' model in all the skill-scores, indicating an improved capability in handling predictive uncertainty through sampling from the probability distribution of the treatment response at the first follow-up, $\hat{Y}_1^{j}$ (see last three rows of Table \ref{tab:skill_scores_longitudinal_T1}). 
The benchmark model based on delta radiomics is more effective in predicting disease progression, with a mean recall of $0.832$ compared to $0.747$ for our G-KDE-based method. However, G-KDE shows a substantial improvement in mean specificity, $0.470$ versus $0.219$ for delta radiomics, showing better ability to predict the stable disease cases, which are severely under-represented at the second follow-up. This results in the enhancement of the balanced accuracy obtained with G-KDE model ($0.609$ vs. $0.526$) and in the increase of the ROC-AUC ($0.639$ vs. $0.601$), see Table \ref{tab:skill_scores_longitudinal_T1}.
G-KDE also enhances balanced accuracy and ROC-AUC compared to 'baseline radiomics' and 'radiomics at $1$st follow-up' models, respectively based on baseline and $1$st follow-up radiomics only. Finally, G-KDE demonstrates comparable performance to the model trained with the true response labels at the first follow-up, 'true labels at $1$st follow-up' model.\\

The proposed longitudinal framework is applied to the specific task of predicting treatment response in glioblastoma, aligning with clinical objectives of outcome prediction in neuro-oncology. The study \cite{benchmark_glioblastoma} provides a reference for radiomics-based prediction of glioblastoma tumor progression after chemoradiotherapy from the baseline timepoint. 
This work included 76 patients, radiomics from contrast-enhanced T$1$, T$2$-weighted, and ADC MRI, clinical characteristics and MGMT promoter methylation status. Outcomes were assessed using baseline radiomics up to six months post-chemoradiotherapy. Key predictive features included age and MGMT promoter methylation status, a known discriminator of overall survival \cite{lumiere}. The model including MGMT achieved a ROC-AUC of $0.80$ with $0.67$ specificity, demonstrating the benefit of combining radiomics with clinical and molecular markers. In contrast, MGMT and age alone reached a ROC-AUC of $0.66$, while radiomics-based models alone presented ROC-AUCs between $0.46$ and $0.69$. 
The performance of our probabilistic model for longitudinal prediction of treatment response aligns with this reference. Moreover, we remark that our longitudinal approach improves the prediction of the baseline model, based on pure baseline radiomics, even with reduced dataset size. This indicates room for improved performance leveraging bigger longitudinal datasets \cite{Bakas2017, Bakas2022, QIN}.

\section{Conclusions}\label{sec:conclusions_longitudinal}
 In this work we introduced a probabilistic model for longitudinal response prediction, 
 where the contribution of baseline features is used to produce a probabilistic estimate of the intermediate outcome, with the final scope of forecasting subsequent responses at successive follow-ups. Specifically, the probability distribution of having response 1 (progressive disease) at the first follow-up was estimated based on the output probabilities produced by a model designed to predict the response at the first follow-up using baseline features. Sample responses were drawn from this distribution and used for testing the longitudinal model, for predicting the response at the second follow-up. This approach leverages the estimated probability distribution of the response at the first follow-up to handle the instrinsic uncertainty of the longitudinal prediction. 

Our approach mitigates a critical limitation of state-of-the-art radiomics-based longitudinal prediction methods, which typically require feature extraction from multiple timepoints. This requirement limits their applicability in prospective settings or in presence of incomplete follow-up data, hindering their ability to effectively handle sparse longitudinal datasets. 
With the proposed framework, by utilizing an ensemble of pre-trained models, the probability distribution of disease progression at the first follow-up could be inferred, thereby enabling a longitudinal assessment at the second follow-up for unlabeled patients, through the pre-trained longitudinal model.

The proposed methodology has been tested using a synthetic dataset and with a real-world radiomics dataset of glioblastoma patients. In this second framework, our longitudinal method is employed for predicting glioblastoma response to treatment at the second post-operative follow-up, using pre-operative radiomics and intermediate response to treatment. 
In the proposed approach, the use of radiomics is restricted to the baseline stage, thereby limiting the growth of problem dimensionality and mitigating the resultant loss of modeling intuition, that would result from indiscriminate aggregation of longitudinal features for model training. 

While the investigation conducted in this study focuses on specific applications, we highlight that the proposed longitudinal method is applicable to various types of imaging features. However, it is particularly suitable for radiomics applications, given that it is designed to address the challenges of longitudinal studies with limited dataset size, and for its architecture, aimed at optimizing the use of features within a context that inherently presents significant constraints and challenges.

Our future endeavors involve the validation and generalization of the proposed methodology, leveraging insights gained from various and larger longitudinal datasets. 
Indeed, a bigger dataset would allow the inclusion of informative clinical biomarkers in the predictive models, like the MGMT, the consideration of an extended radiomic feature set (wavelet and LoG features \cite{wavelet, log}), and a multimodal evaluation.
A larger longitudinal dataset would also enable more comprehensive validation of the methodology and facilitate the establishment of a benchmark for longitudinal prediction models.
Moreover, in this study, the proposed probabilistic framework is used to model the first follow-up response in the test cohort, simulating missing longitudinal data in the target population for model application. Future work will investigate incorporating the probabilistic modeling of intermediate responses into the training phase, thereby removing the dependency on intermediate treatment evaluations during model development. Furthermore, a nested modeling approach will be investigated to enable prediction beyond the second follow-up. Indeed, a highly-accurate well-calibrated model for predicting the first follow-up response would produce reliable probability distribution of disease progression, enabling the training of the longitudinal model with estimated response samples, thus eliminating the need for intermediate treatment evaluation. This would be useful in case of missing intermediate timepoint. Finally, the proposed longitudinal framework should be evaluated in diverse clinical contexts, including neurodegenerative diseases, which inherently require longitudinal analysis.

\end{spacing}
\end{document}